\documentclass[letterpaper]{article}
\usepackage{mume}
\usepackage{times}
\usepackage{helvet}
\usepackage{courier}
\usepackage{musicography}
\usepackage{graphicx} 
\usepackage{float}

\newcommand{\muskern}{\kern-.15ex}
\newcommand{\ppp}{\textnormal{\textbf{\textit{p\muskern p\muskern p}}}}
\newcommand{\fff}{\textnormal{\textbf{\textit{f\muskern f\muskern f}}}}

\title{Music Style Transfer: A Position Paper}
\author{ 
Shuqi Dai\\
Computer Science Department\\
Peking University\\
shuqid.pku@gmail.com\\
\And Zheng Zhang\\
Computer Science Department\\
New York University Shanghai\\
zz@nyu.edu\\
\And Gus G. Xia \\
Computer Science Department\\ 
New York University Shanghai\\
gxia@nyu.edu\\
}
\begin{document} 
\maketitle
\begin{abstract}
\begin{quote}
Led by the success of {\em neural style transfer} on visual arts, there has been a rising trend very recently in the effort of {\em music style transfer}. However, ``music style'' is not yet a well-defined concept from a scientific point of view. The difficulty lies in the intrinsic {\em multi-level} and {\em multi-modal} character of music representation (which is very different from image representation). As a result, depending on their interpretation of ``music style'', current studies under the category of ``music style transfer'', are actually solving completely different problems that belong to a variety of sub-fields of Computer Music. Also, a vanilla end-to-end approach, which aims at dealing with all levels of music representation at once by directly adopting the method of image style transfer, leads to poor results. Thus, we vitally propose a more scientifically-viable definition of music style transfer by breaking it down into precise concepts of timbre style transfer, performance style transfer and composition style transfer, as well as to connect different aspects of music style transfer with existing well-established sub-fields of computer music studies. In addition, we discuss the current limitations of music style modeling and its future directions by drawing spirit from some deep generative models, especially the ones using unsupervised learning and {\em disentanglement} techniques.
\end{quote}
\end{abstract}

\section{Introduction}
\subsection{Background of Automated Music Generation}

The practice of music automation can be traced back to {\em Guido D'Arezzo}, a famous medieval musician who designed a rule-based vowel-to-pitch mapping algorithm to generate a sequence of notes \cite{loy1989composing}. While ``crafting music'' is still the mainstream, {\em algorithmic composition}, or in general {\em automated music generation} has become more and more popular nowadays with the development of modern computers. On the one hand, fast CPUs offer dramatic speedup of experimentations, so that people can test different ideas much more rapidly. In addition, various computer-music programming languages \cite{dannenberg1997machine,mccartney1996supercollider,boulanger2000csound,wang2003chuck} have been invented since the late 1950s¡¯, which further boosted the efficiency of music creation via programming. On the other hand, advanced computational models and data-driven algorithms have empowered computers to generate more human-like music via inheriting certain statistics and styles from the training sets. Recently, with the breakthroughs in artificial neural networks, {\em deep generative models} have become one of the leading techniques for automated music generation \cite{briot2017deep}. For the examples of mimicking J.S. Bach alone, we have seen BachBot \cite{liang2016bachbot}, DeepBach \cite{hadjeres2016deepbach}, CNNBach \cite{huang2016counterpoint}, etc., and most of them can generate convincing results.

Despite these promising progress, people still struggle to generate both {\em natural} and {\em creative} music through automation. In general, algorithms with weak constraints are often ``too random'' and rarely make human-like music, though many works are interesting and creative from a contemporary perspective. On the other hand, algorithms with strong constraints (either explicitly constrained via rules or implicitly constrained by training data) are mostly ``too flat'' and lack the exploration and dynamic that can be easily sensed from genuinely creative works. 

\subsection{Music Style Transfer: Importance \& Challenges}

{\em Image style transfer} techniques \cite{gatys2015neural} inspired the hope to solve the paradox above. By separating and recombining music contents and music styles of different pieces, it is possible to generate new music that is both creative and human-like. In other words, we can still use our favorite data-driven algorithms but twist the constraints or optimizations in general by applying them separately to different aspects (i.e., content and style) of music. 

Such effort is named after {\em music style transfer}. However, there is a severe problem: ``music style'' is a fuzzy term that can literally refer to any aspect of music, ranging from high-level compositional features (such as tonality and chord sequence) to low-level acoustic features (such as sound texture and timbre). This ambiguity is mainly due to the intrinsic {\em multi-level}, {\em multi-modal} character of music representation --- music can be read, listened to, or performed, and it all depends on whether we are relying on {\em score} (the top-level, abstract representation), {\em sound} (the bottom-level, concrete representation), or {\em control} (the intermediate representation). This is very different from image representation, and so far no end-to-end system can deal with all levels of music representation together in an elegant manner. 

Consequently, most studies only focus on a certain level/modality of music representation and therefore have different interpretations of music style. Depending on the interpretation, the essence of music style transfer also varies a lot and may even refer to problems evolved from different sub-fields of computer music, such as algorithmic composition, expressive performance, or sound synthesis. In other words, we are facing an issue of the many-to-one collapse of keyword definition. Without further action, an accumulated upcoming literature all named after ``music style transfer'' would lead to a great confusion of the underlying problems to the readers as well as a risk to ignore the treasures in computer music before the age of deep learning.

In this position paper, we contribute a precise definition of music style transfer based on the uniqueness of music representation. We start from an overview of music representation in Section 2 in order to formally introduce the definition in Section 3, where we also connect different types of music style transfer with existing well-established computer music studies. In the end, we discuss the current limitations and possible future directions of music style modeling by inspecting current unsupervised learning and disentanglement techniques of deep generative models.

\section{Multi-level and Multi-modal Representation}

Music is widely considered a universal language and there are many previous discussions on music representations \cite{dannenberg1993music,wiggins2010non,muller2013freischutz}. The relationship between music notation (score) and actual sound is similar to the one between text and speech. Score serves as a symbolic and highly-abstract visual representation to efficiently record and communicate music ideas, whereas the sound is a set of continuous and concrete signal representations that encode all the details we can hear. Therefore, we can picture the two representations at different levels, with the score at the top and sound at the bottom \cite{dannenberg1993music}. 

In the middle, people often insert an intermediate representation of performance control. The reasons are twofold. First, musical semantics and expression rely heavily on performance control that a funeral hymn can sound really happy by simply tripling the tempo. Second, the performance control for many instruments (e.g., a piano keyboard) can be easily parameterized and therefore very machine friendly. Note that different levels of representation are not solely mutually exclusive, but the multi-level property offers us a useful tool to better understand the essence of music {\em content} and {\em style}. To fully comprehend different aspects of music style transfer, we shall first investigate the multi-level property of music representation more in-depth. 

\subsection{Score Representation}
Score representation exists in many forms, including sheet music notation, lead sheet, chord chart and numbered musical notation. Most of them are highly symbolic and encode abstract music features indicated by the composer, including tonality, chord, pitch, timing, dynamics and rich structure information such as phrases and repetitions. 

The key character of score representation is that the encoded features are mostly {\em discrete} with a mix of measurement scale. Take western music notation (Figure 1) for example. {\em Note onset} is a ratio variable and lies on integer multiples of a certain time unit (usually 1/8 beat is short enough). Pitch is an interval variable, whose corresponding fundamental frequency always lies in a discrete sequence. (E.g., the frequency of C4 in the equal-tempered tuning is 261.63 Hz, the frequency of its successive pitch C\musSharp{}4 is 277.18 Hz, and there is no other pitch frequencies lie in between.) Dynamics is an ordinal variable, usually ranging from \ppp (the softest) to \fff (the loudest). Many other symbols are nominal variables, such as chord types and repeat signs. Such characters bring a challenge for generative models since discrete optimization is in general very difficult and a mixed scale makes some numerical operations impossible. 
\begin{figure}
  \centering
  \includegraphics[width=3.225in, height = 1.4in]{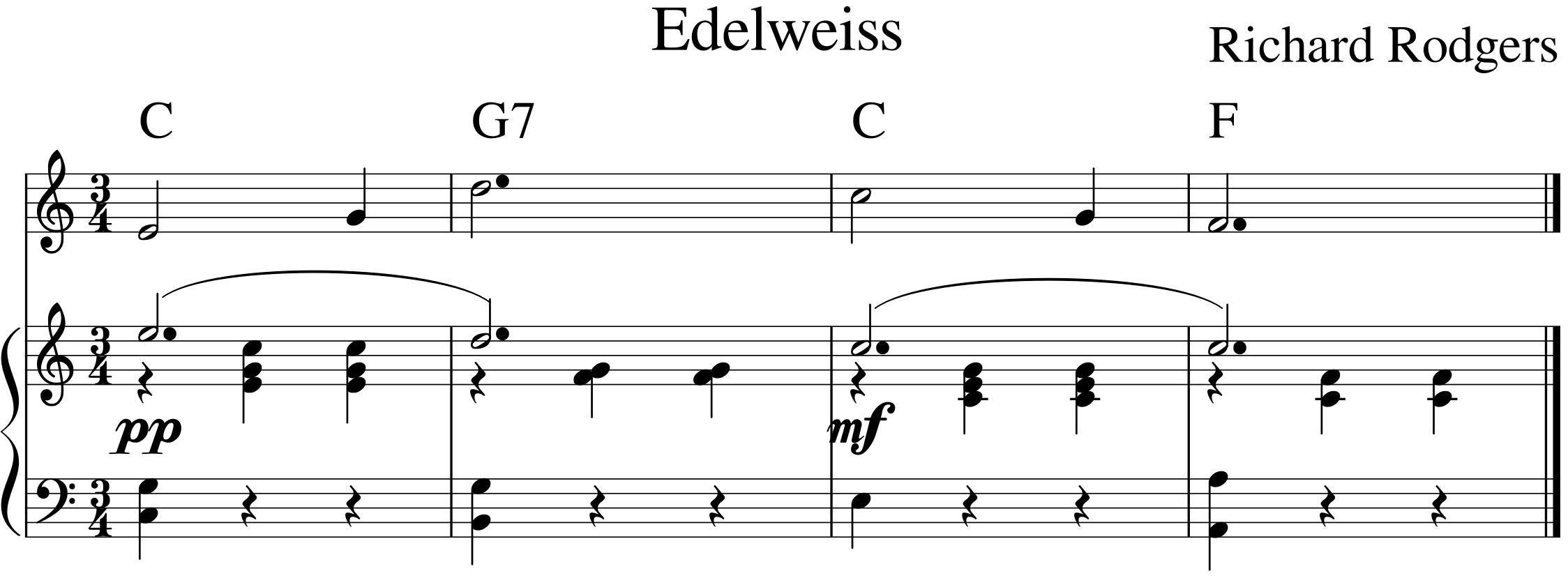}
  \caption{An example of western music notation.}
  \label{1}
\end{figure}

\subsection{Performance Control Representation}
A performance control encodes an interpretation of the corresponding score, rely on which a performer turns the score into performance motions. A commonly used control representation is MIDI piano roll (Figure 2), where each note is encoded by its pitch, dynamics, onset (starting time), and duration. It also has a number of controllers such as pedal and pitch bend for more performance nuances. To be specific, pitches are integers in semitones with C4 being 60, dynamics are integers in velocities units (speed with which the keys are hitting) ranging from 1 to 127, and timings are floating point numbers in seconds. 

Compared to score representation, the key character of performance control is the enriched and detailed timing and dynamics information, which more or less determined the {\em musical expression} of a performance. On the other hand, most structural information such as phrase, repetition, and chord progression is flattened and become implicit during the translation from the score to performance control. Note that performance control is largely independent from the actual instrument; it is not yet the final music sound and still considered a middle-level abstraction.
\begin{figure}
  \centering
  \includegraphics[width=3.125in]{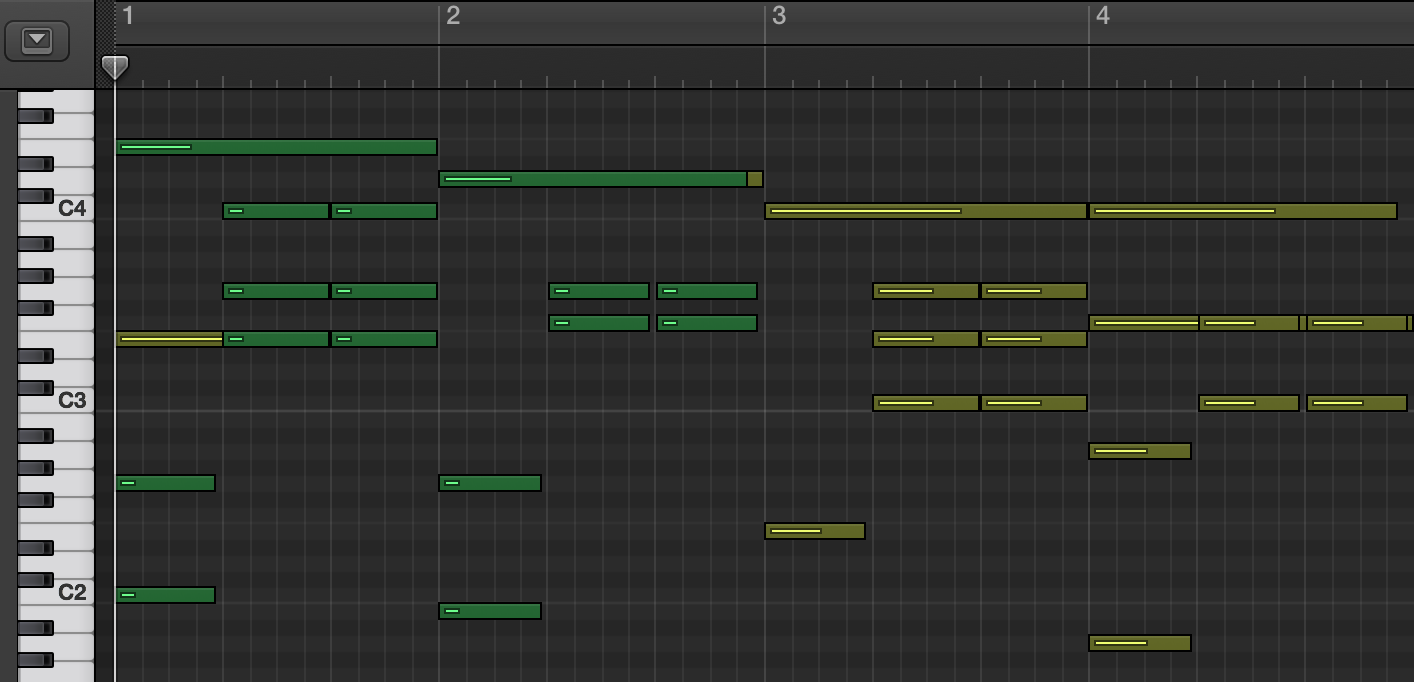}
  \caption{An example of MIDI piano roll representation.}
  \label{2}
\end{figure}

\subsection{Sound Representation}

\begin{figure}
  \centering
  \includegraphics[width=3.125in]{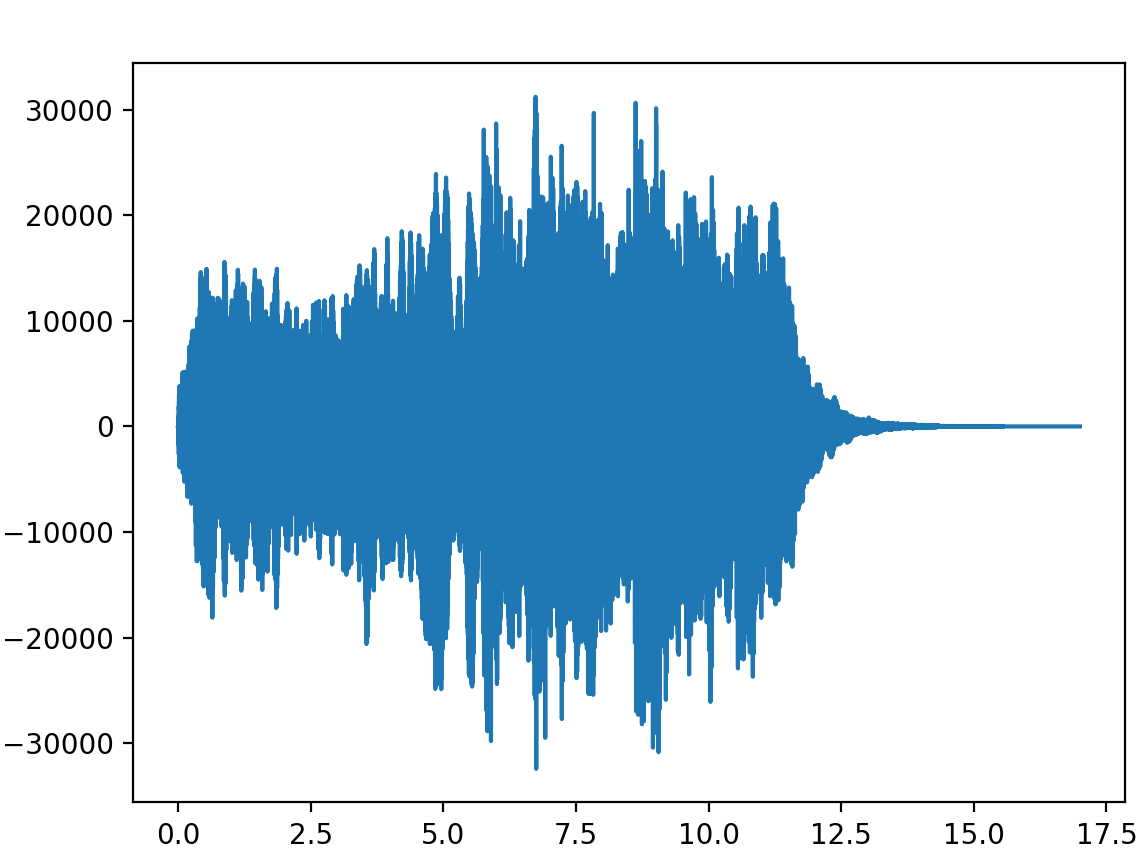}
  \caption{A waveform example where the horizontal axis represents time and the vertical axis represents amplitude.}
  \label{3}
\end{figure}

\begin{figure}
  \centering
  \includegraphics[width=3.125in]{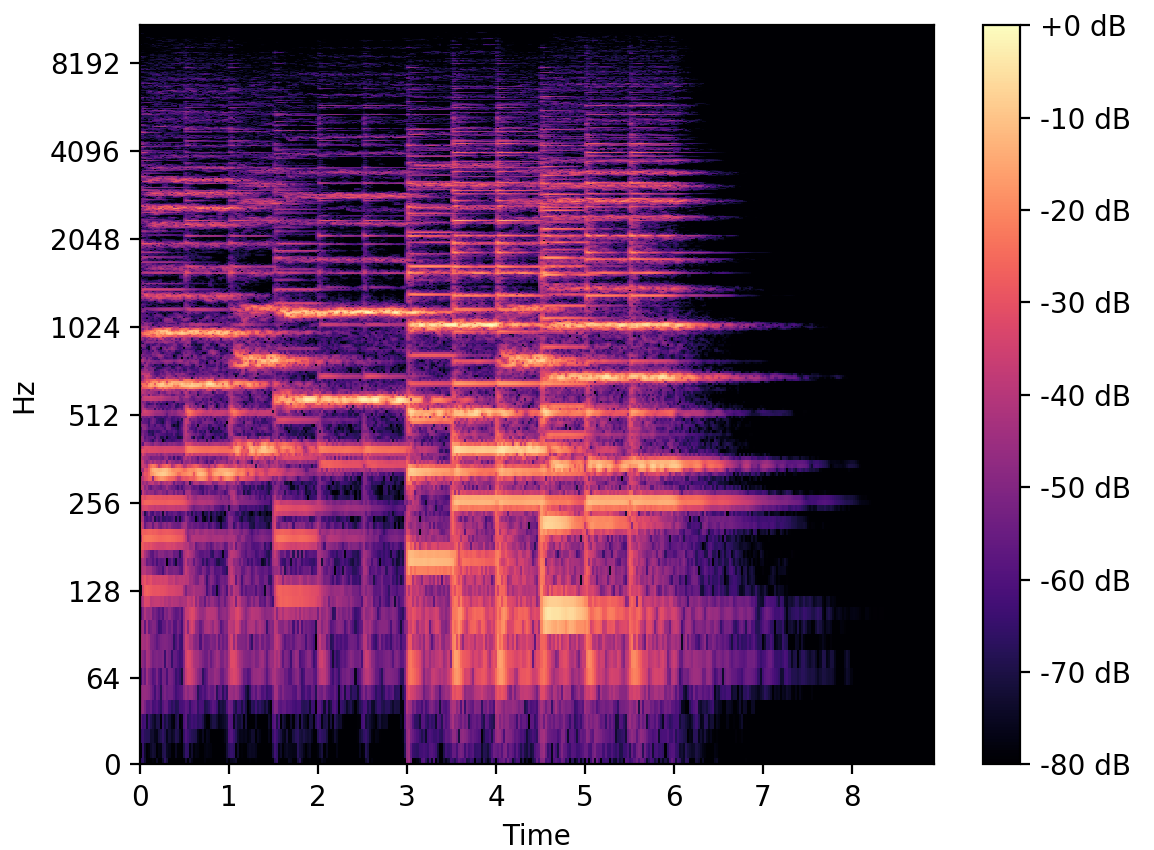}
  \caption{A spectrogram example where the horizontal axis represents time, vertical axis represents frequency, and the color represents energy distribution on different frequencies. }
  \label{4}
\end{figure}

Sound, the concrete signal representation, can be seen as an acoustic realization of the corresponding performance control via a certain instrument. Two commonly used formats for sound representation are waveform (Figure 3) and spectrogram (Figure 4).

The key character of sound representation is purely {\em continuous} and rich in acoustic details such as timbre, articulation, and other nuances not available in other levels of representation. At the expense of such acoustic details, all symbolic abstractions together with precise performance control information become no more explicit and get hidden in the audio.

\subsection{Representation, Content, and Style}

Table 1 shows a summary of different music representations. It is important to notice that the multi-level architecture actually has already implied the essence of music content and music style, i.e., {\em music content is the information extracted through abstraction (from a lower level to a higher level), while music style is the information enriched through interpretation and realization (from a higher level to a lower level). }

\begin{table}[h]
\caption{A summary of music representations.} % title name of the table
\centering % centering table
\begin{tabular}{| l | c | c | c | c |} % creating 10 columns
\hline % inserting double-line
 \ \  & Sensory & Unique & Scale of & Type of \\ [0.5ex]
 \ \  & system & features & measure & data \\ [0.5ex]
\hline % inserts single-line
% Entering 1st row
& & structure \& & &  \\[-1ex]
\raisebox{1.5ex}{Score} & visual & symbolic &  all & discrete \\[-1ex]
\raisebox{1.5ex}{(top) }&  & abstractions &  & \\ [1ex]
\hline
% Entering 2nd row
& & expressive & &  \\[-1ex]
\raisebox{1.5ex}{Control} & motor & timing \& &  \raisebox{1.5ex}{interval} & mixed \\[-1ex]
\raisebox{1.5ex}{(middle) }&  & dynamics & \raisebox{1.5ex}{\& ratio} & \\ [1ex]
\hline
% Entering 3rd row
Sound & & acoustic &  & contin- \\[-1ex]
(bottom) & \raisebox{1.5ex}{auditory}& details & \raisebox{1.5ex}{ratio} & uous\\ [1ex]
% [1ex] adds vertical space
\hline % inserts single-line
\end{tabular}
\label{1}
\end{table}

Thus, a complete end-to-end system for music style transfer should at least fulfill three requirements: 1) be cross-modal and flexible to deal with different measurement scales, 2) automatically extract the performance control and score information from a sound input, and 3) freely manipulate music representations at any level. However, we have to accept the fact that such systems do not yet exist and may not emerge in the near future. The second requirement alone remains an open problem (especially for polyphonic music), and has been the main focus of the whole field of {\em music information retrieval} for many years. 

Therefore, it is beneficial to first solve style transfer for each level of music representation and gradually integrate different components into one system. A hasty attempt at an end-to-end music style transfer system by directly adopting the algorithms for image style transfer \cite{ulyanov2016,gao2017} would only lead to results that sound like a casual remix of different pieces of music.

\clearpage

\section{Music Style Transfer: A Precise Definition \\ and Related Work}

We present the precise definitions of music style transfer for each level of representation in a bottom-up order. They are: 1) {\em timbre style transfer} for sound, 2) {\em performance style transfer} for performance control, and 3) {\em composition style transfer} for score. We also include a brief overview of the related work and connect them with existing sub-fields of computer music.

\subsection{Timbre Style Transfer}
\ \ \ \ {\bf Definition 1:} {\em Timbre style transfer applies to sound representation. It means to alter the timbre information in a meaningful way while preserving the hidden content of performance control.}

A successful timbre style transfer would allow us to reproduce a trumpet performance by a flute with the same musical expression. Timbre style transfer is closely related to {\em sound synthesis} \cite{russ2004sound}, especially the studies aiming to synthesize different sound of acoustic instruments. The difference is that timbre style transfer requires a {\em disentanglement} of timbre (style) and performance control (content) and implies that there is room to create new timbre through the combination of different ones.

Two recent pioneer studies on timbre style transfer are Google's WaveNet autoencoders \cite{engel2017neural} and Stanford's audio spectrograms neural style transfer system \cite{verma2018neural}. The former built an autoencoder for raw waveform using WaveNet (a dilated temporal convolutional neural network). The bottleneck hidden layer is therefore considered a timbre representation and used to create new timbre through linear interpolation. The latter treated audio spectrograms as images and applied image style transfer with additional carefully designed constraints on temporal and frequency energy envelopes.  

We shall also see the limitations. For both works, the disentanglement of timbre and performance control information is not yet very successful, especially when the length of the processed audio unit is long. Also, from a synthesis perspective, the sound quality of synthesized instruments is still far from the state-of-art learning-based synthesis techniques \cite{hu2004automatic} and worth further investigation. As a side note, VisualSoundtrack \cite{ananthabhotla2017visualsoundtrack}, which is named after ``style transfer'', is actually a synthesis system. It requires human inputs of pitch and no disentanglement is involved. 

\subsection{Performance Style Transfer}
\ \ \ \ {\bf Definition 2:} {\em Performance style transfer applies to performance control representation. It means to alter the control information in a meaningful way while preserving the implicit score content.}

A successful performance style transfer would allow us to transfer Louis Armstrong's interpretation of Summertime to the one of Miles Davis. It is closely related to {\em expressive performance rendering}, which studies how to convert static scores into human-like expressive performances by different computational models. \cite{kirke2009survey,widmer2004computational,performance-rnn-2017} The difference is that performance style transfer requires a disentanglement of control (style) and score information (content) and implies that there is room to create new musical expression through the combination of different controls.

As far as we know, there is no work on performance style transfer yet, though performer identification \cite{ramirez2010automatic,stamatatos2005automatic} has been studied for over a decade. One close attempt is the recent Duet Interaction system \cite{xia2016expressive} that can generate an expressive accompaniment based on the performance style of a solo, but it requires a pre-defined score and cannot create new performance styles. As a side note, the work named after ``neural translation of musical style'' \cite{malik2017neural} is actually an expressive performance rendering system, which focuses on dynamic generation given a score input. Thus, performance style transfer remains a brand-new field worth exploring.

\subsection{Composition Style Transfer}
For many forms of score, there is room for further abstraction. Take western music notation for example, the most identifiable score feature, in general, is the melody contour and sometimes with the structural functions of harmony \cite{schoenberg1969structural}. This is especially the case for tonal music.

{\bf Definition 3:} {\em Composition style transfer means to preserve the identifiable melody contour (and the underlying structural functions of harmony) while altering some other score features in a meaningful way.}

A successful composition style transfer would allow us to create {\em variation}, {\em improvisation}, {\em re-harmonization}, or {\em re-arrangement} of a piece of music. A representative masterpiece is the {\em Twelve Variations on ``Ah vous dirai-je, Maman"} by Mozart. Take the first variation for example, it mostly preserved the melody contour and chord progression of the theme and altered the rhythm and texture to a large extent. Recent high-quality pieces (made by human) include: {\em Improvisation of ``Mary had a little lamb''}\footnote{https://youtu.be/Q6Usd3\_fbq8}, a Korean style {\em Carmen Overture}\footnote{https://youtu.be/hKv2\_UCo1ZQ}, and a Chinese style Mozart Sonata\footnote{https://soundcloud.com/wang-michael-452158298/sets/style-transform-of-mozart-sonata}. Composition style transfer is closely related to {\em stylistic automatic composition}, which can be traced back to David Cope \cite{cope1996experiments}. The difference between these two topics is that composition style transfer requires a disentanglement of different score features and implies that there is room to create new types/idioms of score features (such rhythm, texture, and chord progression) through the combination of different ones.

Pioneer studies on automatic composition style transfer include \cite{pati2018,zalkow2016musical,kaliakatsos2017conceptual}, where the first two deal with monophonic composition and the last one deals with polyphonic composition. The work \cite{pati2018} builds pitch and rhythm models separately for different music genres and then create new melodies through the combination of the pitch model of one genre and the rhythm model of another genre. The works by \cite{zalkow2016musical,kaliakatsos2017conceptual} rely on the power of explicit rules to modify melody and merge different chord progressions, respectively. The work \cite{lattner2016imposing} enforces certain music structures by considering additional template-matching constraints in the optimization procedure. 

The demo pieces created by these early studies are still quite immature, especially compared to the pieces made by humans. The major problem is actually not ``how to transfer the composition style'' but ``how to model it'' in the first place. Current composition models still lack the capacity or representation of music structure and the hidden ``grammar'' of chord progressions. Note that most successful cases of the automatic stylistic composition are related to Bach, and at least for non-experts the structure of Bach's compositions is rather local and easy to perceive compared to many other composers. This is unlikely to be a coincidence and worth the attention of future studies.

\section{Future Directions of Music Style Modeling}
How shall we model the styles of composition, performance, and timbre for a better transfer effect? Most current studies use the following three approaches to model music styles: 1) to inherit the style implicitly from the training set \cite{hadjeres2016deepbach,liang2016bachbot,huang2016counterpoint,xia2016expressive}, 2) to use simple style embedding for generation \cite{mao2018deepj}, and 3) to apply style-related constraints for generation. In other words, they all require a manually-defined style representation or style label for generation. 

As stated earlier, {\em style transfer calls for disentanglement of content and style}. It would make more sense to {\em learn} the disentanglement rather than crafting it by hand. In the field of deep generative modeling, learning disentanglement has already attracted a vast amount of attention \cite{thomas2017,karimi2017,larsson2017,kim2017}. For image generation tasks, adversarial training has achieved disentanglement of latent factors and been applied within the generative adversarial network \cite{chen2016infogan} and variational auto-encoder (VAE) \cite{mathieu2016disentangling} framework. A pioneering study has applied the VAE framework for algorithmic composition \cite{roberts2017}. Though the convincing results are still bounded by the length of two bars, it is conceivable to apply it for style transfer task with some modification.

Upon a successful disentanglement, style can be considered as one of the latent factors and style transfer can be accomplished in two steps. The first is to disentangle a ``style" code from the hidden representation that generates the music, and second is to ``plug" such code into an appropriate sequence generation framework that preserves all other factors.

\section{Conclusion}
In conclusion, music style transfer is a new research field which promises novel computational tools to generate both creative and human-like music. Questions like ``what if Miles Davis wrote {\em Twelve Variations on `Ah vous dirai-je, Maman'} and performed it on a flute'' would be no more purely imaginary. In order to generate meaningful results, future works should be aware of the multi-level, multi-modal music representation and be clear whether the focus is timbre style transfer, performance style transfer, or composition style transfer. Also, the automatic disentanglement of content and style representation is the key for high-quality style transfer algorithms and worth the effort from the whole field, and we believe that it is an efficient way, if not the only way, towards a complete end-to-end, cross-modal music style transfer system.

%\section{Acknowledgments}

%\appendix{\LaTeX{} and Word Style Files}\label{stylefiles}

%The \LaTeX{} and Word style files are available on the ICCC-13
%website, {\tt http://computationalcreativity.net/iccc2013/}.
%These style files implement the formatting instructions in this
%document.

%The \LaTeX{} files are {\tt iccc.sty} and {\tt iccc.tex}, and
%the Bib\TeX{} files are {\tt iccc.bst} and {\tt iccc.bib}. The
%\LaTeX{} style file is for version 2e of \LaTeX{}, and the Bib\TeX{}
%style file is for version 0.99c of Bib\TeX{} ({\em not} version
%0.98i).

%The Microsoft Word style file consists of a single template file, {\tt
%iccc.dot}. 

%These Microsoft Word and \LaTeX{} files contain the source of the
%present document and may serve as a formatting sample.  

\bibliographystyle{mume}
\bibliography{mume2018}

\begin{thebibliography}{}

\bibitem[\protect\citeauthoryear{Ananthabhotla and
  Paradiso}{2017}]{ananthabhotla2017visualsoundtrack}
Ananthabhotla, I., and Paradiso, J.~A.
\newblock 2017.
\newblock Visualsoundtrack: An approach to style transfer in the context of
  soundtrack prototyping.
\newblock In {\em International Computer Music Conference (ICMC-2017)}.

\bibitem[\protect\citeauthoryear{Boulanger}{2000}]{boulanger2000csound}
Boulanger, R.~C.
\newblock 2000.
\newblock {\em The Csound book: perspectives in software synthesis, sound
  design, signal processing, and programming}.
\newblock MIT press.

\bibitem[\protect\citeauthoryear{Briot, Hadjeres, and
  Pachet}{2017}]{briot2017deep}
Briot, J.-P.; Hadjeres, G.; and Pachet, F.
\newblock 2017.
\newblock Deep learning techniques for music generation-a survey.
\newblock {\em arXiv preprint arXiv:1709.01620}.

\bibitem[\protect\citeauthoryear{Chen \bgroup et al\mbox.\egroup
  }{2016}]{chen2016infogan}
Chen, X.; Duan, Y.; Houthooft, R.; Schulman, J.; Sutskever, I.; and Abbeel, P.
\newblock 2016.
\newblock Infogan: Interpretable representation learning by information
  maximizing generative adversarial nets.
\newblock In {\em Advances in Neural Information Processing Systems},
  2172--2180.

\bibitem[\protect\citeauthoryear{Cope and Mayer}{1996}]{cope1996experiments}
Cope, D., and Mayer, M.~J.
\newblock 1996.
\newblock {\em Experiments in musical intelligence}, volume~12.
\newblock AR editions Madison, WI.

\bibitem[\protect\citeauthoryear{Dannenberg}{1993}]{dannenberg1993music}
Dannenberg, R.~B.
\newblock 1993.
\newblock Music representation issues, techniques, and systems.
\newblock {\em Computer Music Journal} 17(3):20--30.

\bibitem[\protect\citeauthoryear{Dannenberg}{1997}]{dannenberg1997machine}
Dannenberg, R.~B.
\newblock 1997.
\newblock Machine tongues xix: Nyquist, a language for composition and sound
  synthesis.
\newblock {\em Computer Music Journal} 21(3):50--60.

\bibitem[\protect\citeauthoryear{Dmitry and Vadim}{2016}]{ulyanov2016}
Dmitry, U., and Vadim, L.
\newblock 2016.
\newblock Audio texture synthesis and style transfer.
\newblock
  https://dmitryulyanov.github.io/audio-texture-synthesis-and-style-transfer.

\bibitem[\protect\citeauthoryear{Engel \bgroup et al\mbox.\egroup
  }{2017}]{engel2017neural}
Engel, J.; Resnick, C.; Roberts, A.; Dieleman, S.; Eck, D.; Simonyan, K.; and
  Norouzi, M.
\newblock 2017.
\newblock Neural audio synthesis of musical notes with wavenet autoencoders.
\newblock {\em arXiv preprint arXiv:1704.01279}.

\bibitem[\protect\citeauthoryear{Gao}{2017}]{gao2017}
Gao, Y.
\newblock 2017.
\newblock Towards neural music style transfer.
\newblock Master Thesis, New York University.
\newblock https://github.com/821760408-sp/the-wavenet-pianist.

\bibitem[\protect\citeauthoryear{Gatys, Ecker, and
  Bethge}{2015}]{gatys2015neural}
Gatys, L.~A.; Ecker, A.~S.; and Bethge, M.
\newblock 2015.
\newblock A neural algorithm of artistic style.
\newblock {\em arXiv preprint arXiv:1508.06576}.

\bibitem[\protect\citeauthoryear{Hadjeres and
  Pachet}{2016}]{hadjeres2016deepbach}
Hadjeres, G., and Pachet, F.
\newblock 2016.
\newblock Deepbach: a steerable model for bach chorales generation.
\newblock {\em arXiv preprint arXiv:1612.01010}.

\bibitem[\protect\citeauthoryear{Hu}{2004}]{hu2004automatic}
Hu, N.
\newblock 2004.
\newblock {\em Automatic Construction of Synthetic Musical Instruments and
  Performers}.
\newblock Ph.D. Dissertation, Carnegie Mellon University.

\bibitem[\protect\citeauthoryear{Huang \bgroup et al\mbox.\egroup
  }{2017}]{huang2016counterpoint}
Huang, C.-Z.~A.; Cooijmans, T.; Roberts, A.; Courville, A.; and Eck, D.
\newblock 2017.
\newblock Counterpoint by convolution.
\newblock In {\em 18th International Society for Music Information Retrieval
  Conference (ISMIR-2017)}.

\bibitem[\protect\citeauthoryear{Kaliakatsos-Papakostas \bgroup et
  al\mbox.\egroup }{2017}]{kaliakatsos2017conceptual}
Kaliakatsos-Papakostas, M.; Queiroz, M.; Tsougras, C.; and Cambouropoulos, E.
\newblock 2017.
\newblock Conceptual blending of harmonic spaces for creative melodic
  harmonisation.
\newblock {\em Journal of New Music Research} 46(4):305--328.

\bibitem[\protect\citeauthoryear{Karimi \bgroup et al\mbox.\egroup
  }{2017}]{karimi2017}
Karimi, A.-H.; Banijamali, E.; Wong, A.~W.; and Ghodsi, A.
\newblock 2017.
\newblock Jade: Joint autoencoders for dis-entanglement.
\newblock In {\em Learning Disentangled Representations, NIPS 2017 Workshop}.

\bibitem[\protect\citeauthoryear{Kim and Mnih}{2017}]{kim2017}
Kim, H., and Mnih, A.
\newblock 2017.
\newblock Disentangling by factorising.
\newblock In {\em Learning Disentangled Representations, NIPS 2017 Workshop}.

\bibitem[\protect\citeauthoryear{Kirke and Miranda}{2009}]{kirke2009survey}
Kirke, A., and Miranda, E.~R.
\newblock 2009.
\newblock A survey of computer systems for expressive music performance.
\newblock {\em ACM Computing Surveys (CSUR)} 42(1):3.

\bibitem[\protect\citeauthoryear{Larsson, Nilsson, and
  K\r{a}geb{\"a}ck}{2017}]{larsson2017}
Larsson, M.; Nilsson, A.; and K\r{a}geb{\"a}ck, M.
\newblock 2017.
\newblock Disentangled representations for manipulation of sentiment in text.
\newblock In {\em Learning Disentangled Representations, NIPS 2017 Workshop}.

\bibitem[\protect\citeauthoryear{Lattner, Grachten, and
  Widmer}{2016}]{lattner2016imposing}
Lattner, S.; Grachten, M.; and Widmer, G.
\newblock 2016.
\newblock Imposing higher-level structure in polyphonic music generation using
  convolutional restricted boltzmann machines and constraints.
\newblock {\em arXiv preprint arXiv:1612.04742}.

\bibitem[\protect\citeauthoryear{Liang}{2016}]{liang2016bachbot}
Liang, F.
\newblock 2016.
\newblock Bachbot: Automatic composition in the style of bach chorales.
\newblock Masters thesis, University of Cambridge.

\bibitem[\protect\citeauthoryear{Loy}{1989}]{loy1989composing}
Loy, G.
\newblock 1989.
\newblock Composing with computers: A survey of some compositional formalisms
  and music programming languages.
\newblock In {\em Current directions in computer music research},  291--396.
\newblock MIT Press.

\bibitem[\protect\citeauthoryear{Malik and Ek}{2017}]{malik2017neural}
Malik, I., and Ek, C.~H.
\newblock 2017.
\newblock Neural translation of musical style.
\newblock {\em arXiv preprint arXiv:1708.03535}.

\bibitem[\protect\citeauthoryear{Mao, Shin, and Cottrell}{2018}]{mao2018deepj}
Mao, H.~H.; Shin, T.; and Cottrell, G.~W.
\newblock 2018.
\newblock Deepj: Style-specific music generation.
\newblock {\em arXiv preprint arXiv:1801.00887}.

\bibitem[\protect\citeauthoryear{Mathieu \bgroup et al\mbox.\egroup
  }{2016}]{mathieu2016disentangling}
Mathieu, M.~F.; Zhao, J.~J.; Zhao, J.; Ramesh, A.; Sprechmann, P.; and LeCun,
  Y.
\newblock 2016.
\newblock Disentangling factors of variation in deep representation using
  adversarial training.
\newblock In {\em Advances in Neural Information Processing Systems},
  5040--5048.

\bibitem[\protect\citeauthoryear{McCartney}{1996}]{mccartney1996supercollider}
McCartney, J.
\newblock 1996.
\newblock Supercollider: a new real time synthesis language.

\bibitem[\protect\citeauthoryear{M{\"u}ller \bgroup et al\mbox.\egroup
  }{2013}]{muller2013freischutz}
M{\"u}ller, M.; Pr{\"a}tzlich, T.; Bohl, B.; and Veit, J.
\newblock 2013.
\newblock Freischutz digital: A multimodal scenario for informed music
  processing.
\newblock In {\em Image Analysis for Multimedia Interactive Services (WIAMIS),
  2013 14th International Workshop on},  1--4.
\newblock IEEE.

\bibitem[\protect\citeauthoryear{Pati}{2018}]{pati2018}
Pati, A.
\newblock 2018.
\newblock Neural style transfer for musical melodies.
\newblock Music Informatics Group, Georgia Tech Center for Music Technology.
\newblock https://ashispati.github.io//style-transfer/.

\bibitem[\protect\citeauthoryear{Ramirez, Maestre, and
  Serra}{2010}]{ramirez2010automatic}
Ramirez, R.; Maestre, E.; and Serra, X.
\newblock 2010.
\newblock Automatic performer identification in commercial monophonic jazz
  performances.
\newblock {\em Pattern Recognition Letters} 31(12):1514--1523.

\bibitem[\protect\citeauthoryear{Roberts, Engel, and Eck}{2017}]{roberts2017}
Roberts, A.; Engel, J.; and Eck, D.
\newblock 2017.
\newblock Hierarchical variational autoencoders for music.
\newblock In {\em 31st Conference on Neural Information Processing Systems
  (NIPS 2017) Workshop}.

\bibitem[\protect\citeauthoryear{Russ}{2004}]{russ2004sound}
Russ, M.
\newblock 2004.
\newblock {\em Sound synthesis and sampling}.
\newblock Taylor \& Francis.

\bibitem[\protect\citeauthoryear{Schoenberg and
  Stein}{1969}]{schoenberg1969structural}
Schoenberg, A., and Stein, L.
\newblock 1969.
\newblock {\em Structural functions of harmony}.
\newblock Number 478. WW Norton \& Company.

\bibitem[\protect\citeauthoryear{Simon and Oore}{2017}]{performance-rnn-2017}
Simon, I., and Oore, S.
\newblock 2017.
\newblock Performance rnn: Generating music with expressive timing and
  dynamics.
\newblock https://magenta.tensorflow.org/performance-rnn.

\bibitem[\protect\citeauthoryear{Stamatatos and
  Widmer}{2005}]{stamatatos2005automatic}
Stamatatos, E., and Widmer, G.
\newblock 2005.
\newblock Automatic identification of music performers with learning ensembles.
\newblock {\em Artificial Intelligence} 165(1):37--56.

\bibitem[\protect\citeauthoryear{Thomas \bgroup et al\mbox.\egroup
  }{2017}]{thomas2017}
Thomas, V.; Bengio, E.; Fedus, W.; Pondard, J.; Beaudoin, P.; Larochelle, H.;
  Pineau, J.; Precup, D.; and Bengio, Y.
\newblock 2017.
\newblock Disentangling the independently controllable factors of variation by
  interacting with the world.
\newblock In {\em Learning Disentangled Representations, NIPS 2017 Workshop}.

\bibitem[\protect\citeauthoryear{Verma and Smith}{2018}]{verma2018neural}
Verma, P., and Smith, J.~O.
\newblock 2018.
\newblock Neural style transfer for audio spectograms.
\newblock {\em arXiv preprint arXiv:1801.01589}.

\bibitem[\protect\citeauthoryear{Wang and Cook}{2003}]{wang2003chuck}
Wang, G., and Cook, P.~R.
\newblock 2003.
\newblock Chuck: A concurrent, on-the-fly, audio programming language.
\newblock In {\em International Computer Music Conference (ICMC-2003)}.

\bibitem[\protect\citeauthoryear{Widmer and
  Goebl}{2004}]{widmer2004computational}
Widmer, G., and Goebl, W.
\newblock 2004.
\newblock Computational models of expressive music performance: The state of
  the art.
\newblock {\em Journal of New Music Research} 33(3):203--216.

\bibitem[\protect\citeauthoryear{Wiggins, M{\"u}llensiefen, and
  Pearce}{2010}]{wiggins2010non}
Wiggins, G.~A.; M{\"u}llensiefen, D.; and Pearce, M.~T.
\newblock 2010.
\newblock On the non-existence of music: Why music theory is a figment of the
  imagination.
\newblock {\em Musicae Scientiae} 14(1\_suppl):231--255.

\bibitem[\protect\citeauthoryear{Xia}{2016}]{xia2016expressive}
Xia, G.
\newblock 2016.
\newblock {\em Expressive Collaborative Music Performance via Machine
  Learning}.
\newblock Ph.D. Dissertation, Carnegie Mellon University.

\bibitem[\protect\citeauthoryear{Zalkow}{2016}]{zalkow2016musical}
Zalkow, F.
\newblock 2016.
\newblock {\em Musical Style Modification as an Optimization Problem}.
\newblock Ann Arbor, MI: Michigan Publishing, University of Michigan Library.

\end{thebibliography}

\end{document}